\input epsf
\documentstyle[aps,epsf,preprint,prb]{revtex}
\begin{document}
\title{\bf Prediction of charge separation in GaAs/AlAs
cylindrical Russian Doll nanostructures}
\author{Jeongnim Kim, Lin-Wang Wang and Alex Zunger}
\address{National Renewable Energy Laboratory, Golden Co 80401}
\date{\today}
\maketitle
\def\Ga{\rm Ga}
\def\As{\rm As}
\def\Al{\rm Al}
\def\GaAs{\rm GaAs}
\def\AlAs{\rm AlAs}
\def\Caption#1#2{\vbox{\noindent #2\par}\vskip.1cm}
{\bf
Recent advances in nanotechnology 
permit fabrication of complex nanostructures with special electronic and
optical properties reflecting dimensional confinement
on a nanometer scale,~\cite{NS1,NS2}
{\it e.g.} multiple quantum wells~\cite{MQW}
and core-shell structures.~\cite{QDQW1,QDQW2,QDQW3,QDQW4}
The essential building blocks of such structures are
alternating layers of different semiconducting materials, 
acting as ``wells'' and ``barriers'',
and controlling the confinement energies and,
thus the localization of charge carriers.
Electrons and holes are confined in wells 
and repelled from barriers much like in ``a particle in a box'':
as the well narrows,
the kinetic energy of the confined particle rises.
The materials comprising the
wells and barriers are usually flat,
two-dimensional semiconductor films,\cite{MQW} 
stacked like a deck of cards
to produce ``multiple quantum wells'' or ``superlattices''.
In this case, 
wave functions of the conduction band minimum (CBM)
and valence band maximum (VBM) at the Brillouin zone center,
are localized on
the widest wells, having the lowest confinement energy.~\cite{MQW,FZ95}
We have contrasted the quantum confinement of
(i) multiple quantum wells of flat GaAs and AlAs layers,
{\it i.e.} $(\GaAs)_{m}/(\AlAs)_n/(\GaAs)_p/(\AlAs)_q$, 
with (ii) ``cylindrical Russian Dolls'' --
an equivalent sequence of wells and barriers arranged as concentric 
wires (Fig.~1).
Using a pseudopotential plane-wave calculation,
we identified theoretically a set of numbers ($m,n,p$ and $q$) 
such that charge separation can exist in
``cylindrical Russian Dolls'':
the CBM is localized in the inner GaAs layer,
while the VBM is localized in the outer GaAs layer.
In contrast, the band edge states of 
linear multiple quantum wells
with equivalent layer thickness 
does not exhibit any charge separation,
having equal amplitudes in two GaAs layers, if $m = p$.
Thus, a Russian Doll geometry provides a charge separation that is
impossible with equivalent linear multiple quantum wells. 
%The reason for this is that quantum confinement energies
%are enhanced by {\it curvature}.
This study thus identifies a new geometric degree of freedom (curvature)
that can be used to manipulate electronic properties
of nanostructures.
}

%
%%%%%%%%%%%%%%
In order to avoid approximate ${\bf k}\cdot{\bf p}$ methods
that fail for narrow wells,~\cite{Wood95}
the electronic structure of the nanostructures
is described here using  screened atomic pseudopotentials
in a plane wave basis.~\cite{MZ94} 
Instead of calculating {\it all} eigenstates of the pseudopotential
Hamiltonians (a procedure whose computational cost scales as $N^3$ 
for an $N-$atom system),
we transform the Hamiltonian via the ``folded spectrum method'',
so that only
the physically relevant eigen states around
the band edges are sought and obtained.\cite{WZ94}
The linear scaling of the computational cost of the folded spectrum method 
with system size permits
supercell calculations of rather large, $10^3 \sim 10^4$-atom
nanostructures
needed to study the effect discovered here.

Figures~2 shows the calculated confinement energies of the conduction
band minimum and the valence band maximum  
of linear multiple quantum wells as a function of the thickness 
$p(III_{\Ga})$ of the outer GaAs segment
(see Fig.~1 for definition of the structure).
The confinement energies are defined with respect to
CBM and VBM of the bulk GaAs whose band gap is 1.5 eV.
The innermost GaAs segment is fixed at $m(I_{\rm Ga}) = 5$
monolayers (ML).
We see that, as expected,
both the CBM and VBM are localized on the widest wells.
This is the innermost GaAs segment ($I_{\Ga}$)
when $p(III_{\Ga}) < 2m(I_{\Ga}) = 10$ ML,
and the $III_{\Ga}$ segment 
when $p(III_{\Ga}) > 2m(I_{\Ga})$.
When the two GaAs wells, $I_{\Ga}$ and $III_{\Ga}$,
have the same thickness, $p = 2m$,
the CBM and VBM have equal amplitudes in the two wells
and no charge separation is evident.
The transition in the localization of the CBM and VBM
from $I_{\Ga}$ to $III_{\Ga}$
reflects the dependence of
the confinement energy on
the size of wells, 
as schematically illustrated in Fig.~3.
The confinement energies in well $III_{\Ga}$
increases as the well thickness, $p(III_{\Ga})$, decreases, 
while the confinement energies in well $I_{\Ga}$ remain 
almost constant.
The transition from localization of the CBM and VBM
on $I_{\Ga}$ to localization on $III_{\Ga}$ 
occurs  at $p(III_{\Ga}) < 2m(I_{\Ga})$, 
when the confinement energy of $I_{\Ga}$ 
dips below that of $III_{\Ga}$.
%The reason that in MQW the wave function of the VBM is localized
%in the same spational region as the wave function of the CBM is
%that both act as ``single-band systems''
%($\Gamma_{1c}$ for the CBM and the heavy hole for the CBM),
%without any mixing.

Figure~4 shows the confinement energies of
the CBM and VBM in the {\it cylindrical Russian Dolls}
as a function of $p(III_{\Ga})$;
the thicknesses of other layers are fixed as before.
Similarly to the MQW case of Fig.~2,
both the CBM and VBM are localized in $I_{\Ga}$
when $p(III_{\Ga}) < m(I_{\Ga})$
and in $III_{\Ga}$ when $p(III_{\Ga}) > m(I_{\Ga})$.
However, differently from the MQW,
we observe a charge separation in the wells
for $p(III_{\Ga}) = m(I_{\Ga}) = 10$ ML:
the CBM is localized in $I_{\Ga}$, while the VBM is localized
in $III_{\Ga}$.
We find the same charge separation 
when $p(III_{\Ga}) = m(I_{\Ga}) = 12$ ML,
where the confinement energies are
151.2 meV (CBM) and -30.1 meV (VBM).

The wave functions of the VBM and CBM of the multiple quantum well
and the CBM of the cylindrical Russian Dolls do not change their symmetries
(although their localization can change from $I_{\Ga}$ to $III_{\Ga}$)
as $p$ changes.
Indeed, the CBMs of both structures
are derived from the zincblende $\Gamma_{1c}$ states at all $p$ values,
and the VBM of the MQW is derived from the heavy-hole state at the
Brillouin zone center for all $p$ values.
Since both the VBM and the CBM of the MQW do not change their
identities,
their localization transitions occur at the {\it same} critical
thickness,
so no charge {\it separation} is evident.
In contrast, the VBM of the cylindrical Russian Doll structure 
exhibits, as $p$ increases, a crossing of two levels
with distinct symmetries (circles {\it vs.} triangles in Fig.~4).
Charge separation occurs when the confinement energy
of these two states cross, {\it i.e.} $p = m$.
We emphasize that the charge separation in the cylindrical Russian Dolls
is not due to the band alignment between GaAs and AlAs
(which is the same in Russian Dolls and multiple quantum wells)
but due to the concentric wire geometry 
and the valence band structure.

Table~I gives the confinement energies (insets in Fig.2 and 4)
of the CBM and VBM
for a few structures of cylindrical Russian Dolls and linear multiple
quantum wells.
We see that given the same layer thicknesses, 
the confinement energies ($\Delta E$) of cylindrical Russian Dolls
% --{\it two-dimensionally confined structures} -- 
are considerably larger  than those of 
%{\it one-dimensionally-confined} 
linear multiple quantum wells.
The reason is that the confinement energies
are enhanced by the ``two-dimensional'' nature of the charge carriers
in case of the {\it concentric} layers 
in the cylindrical Russian Doll geometry,
compared to the ``one-dimensional'' nature on the {\it flat} layers
in the linear multiple quantum well structure.
The upper part of Table~I shows
that the confinement dimension together with the well widths
affects the localization of the wave functions,
as shown in Fig.~2 and Fig.~4.
%For $m(I_{\Ga}) = 10$ ML,
%the CBM and VBM of linear multiple quantum wells 
%are localized in wider wells,  {\it i.e.}, 
%in $I_{\Ga}$, since $p(III_{\Ga}) < 2m(I_{\Ga})$.
%Cylindrical Russian Dolls, however, exhibit different localization
%of the band edge states:
%both the CBM and VBM are localized in $I_{\Ga}$ for $p = 4 {\rm ML} < m$,
%while charge separation occurs when $p = m = 10$ ML.

In all cases discussed so far,
all band edge states are $\Gamma-$derived.
However, the bottom half of Table~I
show that when $m(I_{\Ga}) = 6$ ML,
the CBM of cylindrical Russian Dolls is derived
from bulk $X_{1c}$ state and is localized on region $IV_{\Al}$.
Indeed, it has been shown by Franceschetti and Zunger~\cite{FZ95}
that the VBM of the heterostructures consisting of GaAs/AlAs
is always $\Gamma-$ like, while
the CBM becomes $X-$like
as the well width becomes smaller and the confinement increases.
In other words, the CBM is $X-$like,
when the GaAs well is smaller than a critical size.
This transition is found to occur at different critical layer
thickness
in cylindrical Russian Dolls and in multiple quantum wells.
The CBM of the cylindrical Russian Doll changes from $\Gamma$ to $X$-like
when both $m(I_{\Ga})$ and $p(III_{\Ga})$
become smaller than 10 ML.
This critical thickness is consistent with
that for the $\Gamma \rightarrow X$ transition
in an {\it isolated} quantum wire.~\cite{FZ95}
On the other hand, the critical thickness of
the $\Gamma \rightarrow X$ transition in the MQW
is $m = 5$ ML.
Table~I shows therefore that when $m = 6$ ML, 
the CBM of the MQW is $\Gamma-$like,
while that of the cylindrical Russian Dolls is $X-$like.
This illustrate an extreme difference in
electronic properties attainable by different
confining geometries of nanostructures having the same quantum sizes.

In summary,
we have shown that in analogy with nested (Russian Doll)
carbon nanotubes~\cite{Tube},
where new physical properties, absent in the corresponding
flat (graphite) sheets are attainable,
ordinary semiconductor Russian Doll structures
can also exhibit novel properties,
absent in the flat multiple quantum well.
In particular,
Russian Doll GaAs/AlAs structures afford
charge separation on different sheets
and different ($\Gamma$ {\it vs.} $X$) symmetries of states.

%**************************** REFERENCES ****************************

\noindent {\bf Acknowledgements}
This work was supported by
United States Department of Energy -- Basic Energy Sciences, Division
of Materials Science.

%\end{document}
\begin{table}
\caption{Table~I. The confinement energies ($\Delta E$ in meV) of
the CBM and VBM for various layer thicknesses,
$m, n$ and $P$ (in ML) of the cylindrical Russian Dolls and 
multiple quantum wells.
Band edge states are $\Gamma-$like, unless stated.
}
\bigskip
\bigskip
\begin{tabular}{ccrcrc}
Layer thickness&& \multicolumn{2}{c}{Russian Doll$^1$}&
\multicolumn{2}{c}{Quantum Well$^2$}\\
$m-n-p$ & State & $\Delta E$ & Localization & $\Delta E$ & Localization\\
\tableline
10-4-4  & CBM &  181.9 & $I_{\Ga}$ & 84.8 & $I_{\Ga}$\\
       & VBM &  -49.4 & $I_{\Ga}$ &  -20.7 & $I_{\Ga}$\\
10-4-10  & CBM &  170.2 & $I_{\Ga}$ & 83.6 & $I_{\Ga}$\\
       & VBM &  -37.0 & $III_{\Ga}$ & -20.7 & $I_{\Ga}$\\
\tableline
6-4-4  & CBM &  216.1 & $IV_{\Al}$ ($X$) & 167.1 & $I_{\Ga}$\\
       & VBM & -116.7 & $I_{\Ga}$&  -48.7 & $I_{\Ga}$\\
6-4-6  & CBM &  215.8 & $IV_{\Al}$ ($X$) & 165.0 & $I_{\Ga}$\\
       & VBM & -84.5  & $III_{\Ga}$ & -48.7 & $I_{\Ga}$\\
\end{tabular}
$^1$ $q(IV_{\Al}) = 10$ ML and $^2$ $q(IV_{\Al}) = 14$ ML.
\label{CRDvsMQW}
\end{table}

%\end{document}
\newpage
%%%%%%%%%%%FIG. 1
\begin{figure}
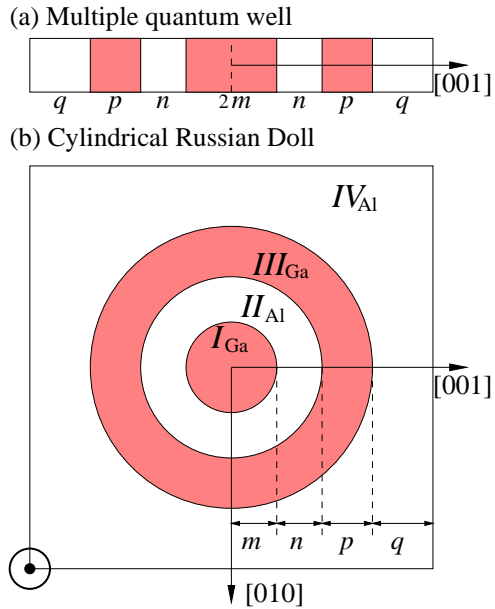

\caption{
Schematics of structure of (a) linear multiple quantum well, whose confinement
direction is indicated by an arrow,
and (b) a $\{100\}$ cross-section of an equivalent
cylindrical Russian Doll.
These structures are made of alternating GaAs (red) and AlAs segments
with thicknesses $m,n,p$ and $q$ monolayers,
in the order (starting from the center)
$I_{\Ga} \rightarrow II_{\Al} \rightarrow III_{\Ga}\rightarrow IV_{\Al}$,
with thicknesses $m, n, p$ and $q$ monolayers, respectively.
}
\end{figure}
\begin{figure}
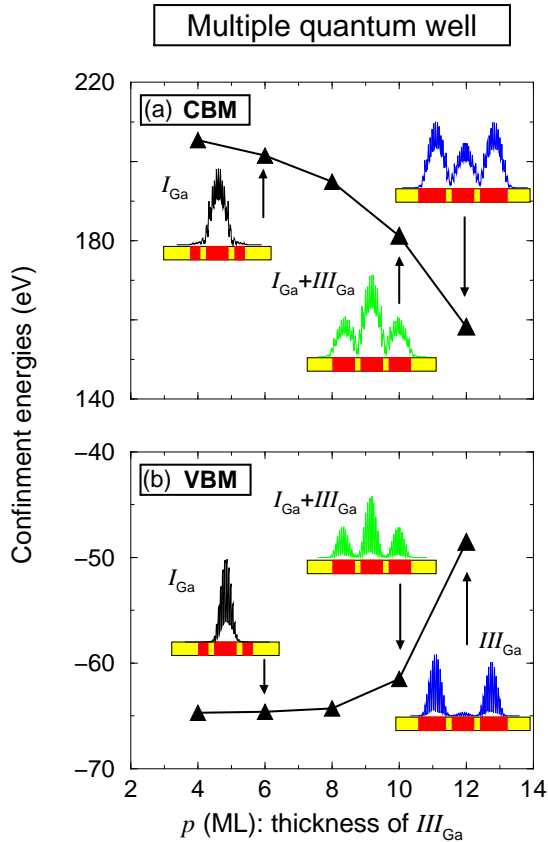

\caption{
Confinement energies (triangles) and wave-function amplitudes (insets)
of the (a) CBM and (b) VBM of linear multiple quantum wells,
as a function of 
the thickness $p(III_{\Ga})$ of the outer GaAs layer.
Other thicknesses are fixed
at $m(I_{\Ga})= 5$ ML, $n (II_{\Al})= 4$ ML and
$q(IV_{\Al}) = 8$ ML.
Note that the CBM and VBM are always localized on the {\it widest} wells:
on $I_{\Ga}$ for small $p$, and on $III_{\Ga}$ for large p.
}
\end{figure}
\begin{figure}
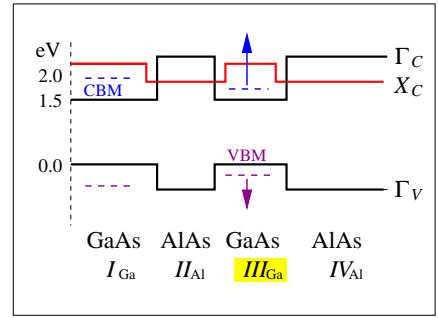

\caption{
Band alignment of the GaAs and AlAs layers 
along the confinement direction for the MQW
and along the radial direction for the cylindrical Russian Doll (see Fig.~1).
The arrows indicate the movement of confined levels
as the size $p(III_{\Ga})$ decreases,
while the thicknesses of other layers are held fixed.
In a conventional linear multiple quantum well,
both the CBM and VBM levels are localized
on the widest well, having the lowest kinetic energy confinement,
thus the lowest energy levels in the respective wells (Fig.~2).
For the same well thicknesses ($m = p$),
the band edge states have similar amplitude on $I_{\Ga}$ and $III_{\Ga}$.
In contrast,
in cylindrical Russian Dolls (Fig.~4),
we can have the VBM on region $I_{\Ga}$,
while the CBM is localized in region $III_{\Ga}$,
even though  $m = p$.
}
\end{figure}
\begin{figure}
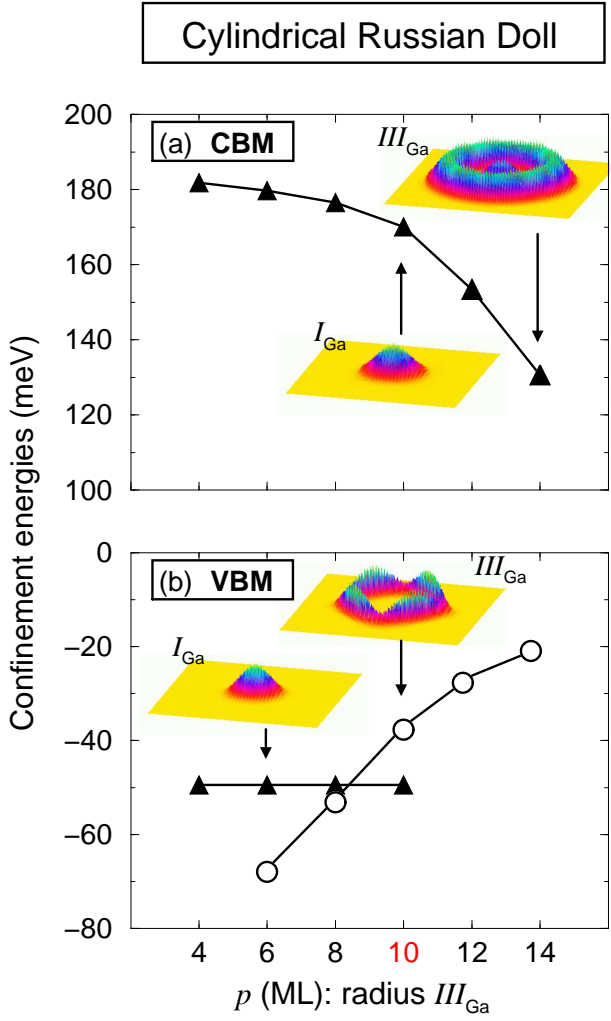

\caption{
Confinement energies of the (a) CBM and (b) two highest valence bands
for cylindrical Russian Dolls {\it vs.} the thickness $p(III_{\Ga})$.
The other parameters are held fixed at
$m = 10$ ML, $n = 4$ ML and $q = 8$ ML.
Wave-function amplitudes, averaged along the wire direction,
are shown as insets for a few structures.
Note the change in localization of the wave functions from $I_{\Ga}$
to $III_{\Ga}$.
A charge separation of the electron and hole in the GaAs wells
is obtained at $m (I_{\Ga}) = p(III_{\Ga}) =10$ ML
when the level crossing and attendant change in
angular symmetry of the VBM wave functions occur;
there is no change in angular symmetry of the CBM.
}
\end{figure}

\end{document}